\documentclass[preprint,amsmath,amssymb,amssymb,superscriptaddress,aps,pra,nonbibnotes]{revtex4-1}
\usepackage[colorinlistoftodos, color=green!40, prependcaption]{todonotes}
\usepackage{graphicx,url}
\usepackage[utf8]{inputenc}  
\usepackage[english]{babel}
\usepackage{physics}
\usepackage{float}
\usepackage{url}
\usepackage{soul}
\usepackage{graphicx}
\usepackage{dcolumn}
\usepackage{bm}
\usepackage{times}
\usepackage{mathrsfs}
\usepackage{color}
\usepackage{todonotes}
\usepackage{natbib}
\usepackage{algorithm}
\usepackage{algorithmic}
\begin{document}
	\title{ A Cyclical Deep Learning Based Framework For Simultaneous  Inverse and Forward design of Nanophotonic Metasurfaces.}
	\author{Abhishek Mall}
	\affiliation{Department of Physics, Indian Institute of Technology -- Bombay, Mumbai - 400076, India}
	
	\author{Abhijeet Patil}
	\affiliation{Department of Electrical Engineering, Indian Institute of Technology -- Bombay, Mumbai - 400076, India}
	
	\author{Amit Sethi}
	\email{asethi@iitb.ac.in}
	\affiliation{Department of Electrical Engineering, Indian Institute of Technology -- Bombay, Mumbai - 400076, India}
	
	\author{Anshuman Kumar}
	\email{anshuman.kumar@iitb.ac.in}
	\affiliation{Department of Physics, Indian Institute of Technology -- Bombay, Mumbai - 400076, India}

	
	\begin{abstract}
		The conventional approach to nanophotonic metasurface design and optimization for a targeted electromagnetic response involves exploring large geometry and material spaces, which is computationally costly, time consuming and a highly iterative process based on trial and error. Moreover, the non-uniqueness of structural designs and high non-linearity between electromagnetic response and design makes this problem challenging. To model this non-intuitive relationship between electromagnetic response and metasurface structural design as a probability distribution in the design space, we introduce a cyclical deep learning (DL) based framework for inverse design of nanophotonic metasurfaces. The proposed framework performs inverse design and optimization mechanism for the generation of meta-atoms and meta-molecules as metasurface units based on DL  models and genetic algorithm. The framework includes consecutive DL models that emulate both numerical electromagnetic simulation and iterative processes of optimization, and generate optimized structural designs while simultaneously performing forward and inverse design tasks. A selection and evaluation of generated structural designs is performed by the genetic algorithm to construct a desired optical response and design space that mimics real world responses. Importantly, our cyclical generation framework also explores the space of new metasurface topologies. As an example application of utility of our proposed architecture, we demonstrate the inverse design of gap-plasmon based half-wave plate metasurface for user-defined optical response. Our proposed technique can be easily generalized for designing nanophtonic metasurfaces  for a wide range of targeted optical response.
		
	\end{abstract}
	\maketitle

	
	\section{Introduction}

	The design and optimization of metasurfaces for unconventional functionalities has led to several new optical meta-devices for control over propagation of electromagnetic (EM) waves at sub-wavelength scale  \cite{jin2020experimental, wen2020multifunctional, zhou2020switchable, fan2020reconfigurable, gomez2020enhanced}. The constituent nanostructures of a metasurface are typically meta-atoms and meta-molecules which are designed for various light modulation applications  \cite{zhang2020photonic,ruffato2019total}. Conventional methodologies of design and optimization of metasurfaces components are iterative, trial and error based solvers, and rely on physics-inspired approaches. These approaches employ numerical full-wave simulations such as finite-difference time-domain (FDTD) or finite-element method (FEM) for solving Maxwell's EM equations. However, the development of optimized metasurfaces for various functionalities requires one to go beyond the limitations of physical intuition. Recently, DL has been used for inverse design and optimization of metasurface-based nanostructures for directed functionality \cite{gao2019bidirectional, ma2019probabilistic,ma2018deep,so2019designing,an2020freeform,jiang2019free}. With the use of multiple models based on DL, computational expense has been significantly reduced, and design and optimization processes have become highly efficient.
	\\
	Unlike conventional design approaches, DL is a data-driven method that uses convex optimization to fit a layered non-linear model from inputs to the desired outputs. DL framework is composed of hierarchical artificial neural networks (NNs), known for their ability to learn the highly non-linear patterns in the training dataset. With a sufficient amount of training data and regularization techniques, the learned representation from NNs can generalize to unseen datasets. Conventional fully connected (FC) NNs \cite{beale1996neural} and convolutional neural networks (CNNs)  \cite{fukushima1989analysis} have been used for nanophotonic metasurface design and optimization for the targeted optical response \cite{an2019deep}. Most of these methods either encounter the limitation of optimizing a single candidate design or the requirement of a large dataset for the training process. NNs have also been used as a cascaded \cite{liu2018training} architecture with forward and inverse design networks for several targeted functionalities in nanophotonic metasurfaces. This NN architecture addresses the inverse design of nanophotonic metasurface as a regression problem, mapping optical response to structural design space. This approach however, forces the network converge to one of the several solutions. These DL methods lack flexibility in designing nanophotonic structures because they usually limit the process of optimization to a predefined design of candidates and cannot generate new metasurface designs. \\
	
	Few studies  have tried to formulate the inverse design problem as modeling a conditional probability distribution of geometry/design for a given optical response  \cite{tang2020generative, hodge2019joint} using  a conditional generative adversarial network (cGAN)  \cite{mirza2014conditional} and variational autoencoders (VAEs)  \cite{pu2016variational}. In addition, global optimization techniques such as genetic algorithms (GAs) have also proved useful for inverse nanophotonic design \cite{382334, jin2019complex}. Nevertheless, generative approaches can often lead to structural designs with a deviating optical response and may require longer training and a larger dataset to generate highly efficient structural designs \cite{ma2019probabilistic}. On the other hand, GAs  face the issue of poor generalization capability to parameter space of different topologies \cite{elsawy2019global}. Recently, simultaneous training of NNs performing forward and inverse mapping has shown promising results \cite{liu2018generative,liu2020hybrid}. These approaches use generative models for inverse mapping and CNNs with multiple layers for the forward mapping. However, simultaneously training these NNs introduces problems in proper hyper-parameters selection due co-dependence of both NNs and can result in poor convergence \cite{liu2018generative}.
	
	In this paper,  we propose a cyclical generation and discovery of metasurfaces for user-defined optical response, with no structural design restrictions during inverse design. Our framework consists of efficiently trained deep generative model and a simulation neural network (SNN) - performing inverse and forward design, respectively guided by a pseudo genetic algorithm (pGA) scheme. We use a conditional generative adversarial network (cGAN) as a generative model to model the probabilistic distribution of design space and generate new structural designs while using the SNN to evaluate the authenticity of the structural designs generated by predicting the corresponding optical response. The pGA is applied to the optical responses to distinguish designs with minimal variance in their optical response compared to the desired optical response and sort them to create a desired optical response and design space through selection and evaluation process. The proposed framework works cyclically to generate a desired optical response and design space that mitigates the need to collect data from traditional numerical simulation methods to train and analyze DL models. The optimized framework generates structural designs with 0.021 mean square error (MSE) and 0.968 cosine similarity, while also discovering new metasurface designs.
	\section{Methodology }
	
	\subsection{Dataset Preparation}
	
	For the training and evaluation of our DL-NNs, we generate a dataset of 1500 Aluminum (Al)-nanoantennae samples as meta-atom and meta-molecule units\cite{wu2017versatile} of a periodic gap plasmon based half-wave plate metasurface (HM) with four classes of structural design (rectangle, double-arc, rectangle-circle pair, rectangle-square pair) of different dimensions. A typical sample from the dataset is shown in Fig.~1(a), which is a pair of Al-nanoantennae structural design represented as a 2D-cross sectional image of 64$\times$64 pixels and with corresponding conversion efficiency of an incident left circular polarization (LCP) to a right circular polarization (RCP) as optical response. The unit cell size of 230~nm x 230~nm with different structural design of antennae  consists of Al-nanoantennae of thickness 50nm placed on a SiO$_2$ dielectric spacer over a 150nm thick Al-reflector on a silicon substrate (see Fig.~1(a)). The dielectric spacer thickness significantly affects the conversion efficiency, hence we encode the pixel intensity of structural designs to represent the dielectric spacer thickness $(t)$ in range of 50-150 nm. The reflection conversion efficiency spectrum of each HM with 101 spectral points in wavelength range of 400 nm to 800 nm is obtained by FEM-based electromagnetic simulation software: COMSOL MULTIPHYSICS with Livelink for MATLAB.
	
	From the total dataset, 75\% is used for the training phase and the remaining 25\% is used to test the trained DL models.  As shown in Fig.~1(b), the data set with different structural designs of varying dimensions is used for the training of DL models.  The DL models used solves metasurface design and optimization problem for a fixed unit cell size, periodicity, and wavelength range. Nonetheless, this design and optimization is not limited due to these assumptions and can be generalized by adding more design groups for nanoantennae with variable dimensions that more strongly influence the optical response. In addition, the proposed DL models and the cyclic generation framework can be used to incorporate additional geometrical or material parameters by encoding the information as pixel intensity in the image or concatenated at input node\cite{an2020freeform}.
	
	
	\subsection{Configuration and Training of DL models}

	In the next step, we implement the DL models using PyTorch libraries with hyper-parameters of the training process as mentioned in  Table III in the Supplementary Information. In the proposed framework, the two DL models act as forward design and inverse design network.
	\\
	
	\textbf{The Forward Design Network : } We implement an optical response predicting forward  design neural network - Simulation Neural Network (SNN), as a combination of CNNs and FC-NNs architecture. The target of the SNN is to model the underlying non-linear relationships between the structural design of Al-nanoantennae and the corresponding optical response as a forward one-to-one mapping. Fig.~2(a) shows the SNN model architecture which performs the task of extracting features using CNNs and forward mapping via FC-NNs from the 2D cross sectional image as input and prediction of the corresponding optical response as output. As shown in Fig. 2(a), the convolution section (blue dashed box) of the SNN consists of three blocks each with two convolutional layers followed by a max-pooling layer.  The flattened features from the convolution section are fed to a fully connected network (orange dashed box) with four FC layers. Batch normalization is used after each layer in the proposed architecture. Rectified linear unit (ReLU) is used as non-linearity in each layer except the output layer. The output layer is followed by a sigmoid function. The SNN's detailed architecture is given in Table I of the Supplementary Information. \\
	SNN is trained using the standard backpropagation algorithm to optimize mean square error (MSE) between the predicted ($R^\prime$) and actual optical response ($R$) given in equation (1) as: 
	
	\begin{equation}
	L_{MSE}=  \frac{1}{N}\sum_{i=1}^{N}(R' -R)^{2} 
	\end{equation}
	
	where $N$ is a mini-batch of training dataset. We used regularization methods of dropout and weight decay to avoid over-fitting. An efficiently trained SNN acts as a numerical EM simulation solver and in a few milliseconds predicts an optical response for a structural design input as an image.

	\textbf{The Inverse Design Network : } Here we use DL based conditional generative adversarial network (cGAN)  to generate structural designs of meta-atom and meta-molecule as a unit of HMs for a user-defined optical response. The network produces Al-nanoantennae structural designs in the form of 2D cross sectional images for conversion efficiency as a desired optical response. The cGAN model is based on CNNs which use multiple feature maps to perform convolution operations to extract hidden features as input data from images.\\
	The cGAN  model architecture is shown in Fig.~2(b), consisting of two CNN-based networks: a generator $(G)$ and a discriminator $(D)$ -- aiming to train the generator NN to generate good quality Al-nanoantennae structural designs for a given input optical  response. In our cGAN model, $G$ consists of five transposed CNNs while $D$ consists of five CNNs and two FC-NNs. The detailed architecture is described in Table II of the Supplementary Information. The $generator$ takes 512x1 dimensional input as concatenation of random noise and vector representation of optical response (R). We have considered 101 spectral points for optical response. During cGAN model training, $G$ learns about the conditional probability distribution of Al-nanoantennae structural design space and produces a 2D cross sectional image given an optical response as input.  On the other hand, $discriminator$ takes the structural design, $G(z)$, generated from $G$, the real structural designs $x$ from training dataset and condition with optical response as inputs and identifies them as real (actual designs) or fake (generated designs by $G$).  Essentially, both $G$ and $D$ are trained simultaneously until $G$ learns to generate near real structural designs to deceive $D$. During the training, $D$ tries to identify fake images through minimizing the classification error while $G$ tries to generate inputs so that $D$ can not distinguish between real and fake images. $G$ and $D$ are trained to minimize and maximize the cost function respectively given by:
	
	\begin{equation}
	\min_{G}\max_{D}\ L(D,G) =  E_{x\sim \textrm{P$_{data}$}(x)} \log [ D(x|R)]  + E_{z\sim \textrm{P$_z$}(z)} \log [1 - D(G(z|R))] 
	\end{equation}
	
	where $D(x|R)$ represents the probability of structural design being real from training dataset  for given input optical response, and $D(G(z|R))$ is probability of structural design generated by $G$ given the input optical response. In the cGAN model, $D$ is trained to maximize expectation value of $E$ with  $ E_{x\sim \textrm{Pdata}(x)} \log [D(x|R)]$ for a real structural design image and $E_{z\sim \textrm{Pz}(z)} \log [1 - D(G(z|R))] $ for image generated by $G$. Whereas $G$ is trained to give minimized expectation values to fool $D$. This adversarial training lets $G$ to generate high quality 2D cross sectional image of Al-nanoantennae structural designs. \\
	The training phase of SNN and cGAN converged in 500 and 2000 training steps, respectively using a single 4 GB Nvidia GeForce GTX 1050 GPU where each training step takes less than one minute. The training for SNN was terminated when there was no further progress in the model's accuracy on the validation data set, and cGAN produced good structural design efficiently.

	\subsection{Evaluation of DL models}
	
	For the evaluation of trained DL models, we measure the accuracy of the predicted optical response by SNN and authenticity of optical response of generated designs through cGAN, using two metric:  $MSE$ and cosine  similarity. The MSE evaluates the average error of the optical response per spectral point given as in Eq.~1. Whereas, we calculate the similarity between optical responses as the cosine similarity\cite{dehak2010cosine} formulation as given in equation (3) :
	
	\begin{equation}
	\text{Similarity} = \cos{\theta} = \frac{\sum_{i=1}^{n} R_i R_i'}{\sqrt{\sum_{i=1}^{n}  R_i^2}.\sqrt{\sum_{i=1}^{n} {R'}_i^2}}
	\end{equation}
	where $R$ and $R^\prime$ are real optical response and predicted optical response respectively. The cosine similarity of 1 reflects identical characteristics between two optical responses.
	
	Here we use the metric of cosine similarity to measure the similarity between the two optical responses. This choice of this metric is due to the presence of sharp variations in the optical response in the EM simulation dataset. These sharp variations include combinations of dips/peaks, oscillations and flat reflections (see Fig.~1 of Supplementary Information). In addition, the optical responses for the training process were simulated for randomly generated structural design parameters. There may be cases where certain spectral features occur in fewer instances of the data sample thereby making it difficult for DL models to learn those features\cite{wiecha2019deep}. Therefore, for two optical responses as vectors in response space, the MSE measures the difference between the vectors while cosine similarity measures the similarity of features to ensure accurate prediction of optical response by SNN.
	
	\subsection{The Cyclic Generation Framework}
	
	Using the above mentioned models based on DL, we set up an optimization routine for Al-nanoantennae structural design for user-defined optical response. Here we use a pseudo genetic algorithm (pGA) to search for optimized and authentic structural designs using an objective function from designs generated from DL models. Fig.~3 demonstrates the inverse design framework using both the DL models, cGAN and SNN followed by pGA.
	
	The design procedure begins with initializing a batch of $M$ user-defined desired optical response as a Gaussian mixture. The Gaussian mixture is given by:
	\begin{equation}
	g(\lambda)= \sum_m g_m \exp\{-\frac{(\lambda - \lambda_m)^2}{2\sigma_m^2} \},
	\end{equation}
	where $m$, $\lambda_m$ and $\lambda$ are the number of Gaussian, central wavelength of $m^{\text{th}}$ Gaussian and wavelength range respectively.
	Each sample in the batch is a variety of mixture of gaussian introduced with random shifts in $\lambda_m$. After initialization, the batch of desired optical response ($R^d$) is input into the cGAN model to generate corresponding structural designs. Subsequently, to obtain the predicted optical response ($R^\prime$), the structural designs generated are given input into the SNN model. We now use pseudo genetic algorithm to select the structural designs having optical response best optimized and generated in accordance with the desired optical response as input. The pGA first selects and evaluates the samples generated on the basis of the measure of their MSE and then cosine similarity.\\ 
	In typical genetic algorithm technique, the problem of optimizing complex metasurface designs often involves minimization/maximization of objective or fitness function to lead towards convergence. Instead, here we evaluate the optical response of generated structural designs to guide the process of design and optimization by evaluating each sample firstly on the basis of its minimum MSE and then maximum cosine similarity score. Therefore our implementation of the genetic algorithm is addressed as `pseudo', where we sample from the generated structural designs for its optical response deviation from desired optical response instead of sampling from the gaussian and mutating its mean and standard deviation.  We control the MSE objective function minimization by specifying a threshold value ($v$) for each batch of input optical response. This threshold incorporates maximum error propagation from both the DL models. The pGA selects $M_{\text{best}}$ desired optical response and generated structural design pairs from $M$ initialized batch samples firstly for minimum MSE  and then with maximum cosine similarity and sorts them in the desired response and design space. This characterized distribution of desired response and design space acts as new generation and along with EM simulation dataset (parent) is used as an updated training set for cGAN and SNN to move further for model optimization and learn the latest data representation and correlation for desired response and generated designs (see Fig. 3). The new training dataset (next generation) is, after each cycle step, a combination of newly generated desired response and design space (new individuals) and data set used in previous stage training (parents). The entire DL and pGA framework performs a cyclic process generation, simulation, selection and evaluation to create a desired response space with optimized structural designs and accurate optical responses; hence performing inverse and forward design simultaneously. In \textbf{Algorithm 1}, we provide a simple illustration of the cyclic generation framework.
	
	\begin{algorithm}[H]
		\caption{Steps of cyclic generation}
		\begin{algorithmic}[1]
			\label{alg:the_alg}
			\STATE Start with batch of $M$ user-defined gaussian shape optical response.
			\STATE Generate structural designs and predict the corresponding optical response using cGAN model and SNN model respectively.
			\STATE Compute the objective function MSE and cosine similarity for all samples of batch for given threshold ($v$) of MSE.
			\STATE Sort the $M_{\text{best}}$ generated samples for optimized designs (firstly with minimum MSE and then maximum cosine similarity).
			\STATE Update for training with desired optical response and design space along with previous* dataset.\STATE Repeat steps (2-5) until cGAN and SNN training converges to better optimal.
			
			\textit{*previous dataset for first cycle is EM simulation}
		\end{algorithmic}
	\end{algorithm}
	
	After each cyclic step, $M_{\text{best}}$ designs and response pair obtained are considered as observations and updated in training dataset for next cyclic step. In each step whole framework is optimized by learning updated design and response space correlations. Because the updated desired response and design space have data samples that mimic the real world user-defined responses, learning such new correlations as probabilistic data distribution could make simpler for the cGAN and SNN network to generate new designs and predict more accurate response respectively. In addition, our approach is based on cyclic generation of a design and response space and updating training dataset, hence the associated DL models seek for more optimal convergence and model accuracy improvement.

	\section{Results and Discussion}
	
	Once the training process for DL models is complete, we evaluate and test the effectiveness of the SNN and cGAN for optical response prediction and structural design generation, respectively by employing them on unseen dataset samples.
	
	For SNN's forward design functionality, we test it against randomly selected Al-nanoantennae structural designs from each class to visualize SNN 's predictability. The trained SNN can predict conversion efficiency for the unseen structural designs, with average MSE of 0.026 and cosine similarity of 0.954.  Four such samples of data are shown in Fig. 4, demonstrating the accuracy of SNN on different classes of structural designs, where the real optical response from EM simulation (red curve) and SNN’s predicted (grey circles) optical response are in good agreement. Also, prediction of optical responses using SNN takes a few milliseconds of time. These results show that SNN can be used as a proxy for the time-consuming process of the forward mapping. We also note that at some sharp resonances the SNN model was not able to approximate spectra much precisely. It may be that the model's MSE (loss function) is averaged over all spectral points, and thus the sharp resonance error is diminished as a single spectral point in total MSE, however the high cosine similarity helps in  mapping the spectral features of optical responses efficiently. A separate training of such a NN for a forward design task is more reliable because integrating a NN into the training along with inverse design NN introduces oscillating errors due to mismatch in prediction of optical response for  generated structural designs and real optical response. Simultaneously, the whole NN framework gets trapped in local minima and works less effectively \cite{liu2018generative}. Our SNN learns pixel representation of structural designs and correctly models the strong mapping to its EM simulation.

	For the inverse design process, when the training is complete, cGAN can produce a structural design for an optical response as input in less than a second. Fig.~5 shows randomly selected test results from each structural design class. The real structural designs (black box) and the corresponding structural designs generated by cGAN (red box) show good qualitative agreement. The EM simulations spectra for real structural designs (black curve) and generated structural designs (red curve) along with SNN predicted spectra for generated structural designs show excellent quantitative accuracy. For generated and real structural designs, the average MSE and cosine similarity over the test samples is 0.011 and 0.987, respectively which highlights the potential of using the trained cGAN architecture to generate structural designs for user-defined optical response.

	We test the performance of our framework which includes above trained SNN, cGAN and followed by pGA by inverse designing the gap-plasmon-based half-wave plate metasurface with optical responses for the desired conversion efficiency. The algorithm runs for 5 cycles with a  batch initialization of 1000 samples of desired gaussian mixtures. The threshold defined on the pGA is MSE 0.037 which is the sum of the SNN and cGAN average MSE error considering the cumulative propagation of errors from DL models. After 5 cycles of DL-pGA framework, the on-demand retrieval process results in Fig.~6 showing a few samples from the desired response and design space. Fig.~6(a)-(d) show the optical response predicted by the SNN for the generated structural designs for a desired optical input response from each class. Comparing the two optical responses, it is clear that our proposed framework successfully  generates designs for the desired optical response with 0.021 MSE and 0.968 cosine similarity and replicates it with only minor deviations. Our framework is robust in generating structural designs with optical response closest to the desired optical response, although there may be a possibility of design non-existence. However, after 5 cyclic frame runs, we observe the framework's learning ability to generalize a to a new structural design as shown in Fig.~7(a)-(b) where we see new structural design as either intermediate design (an axe shape) from training designs or double arc design evolving into a dumbbell shape respectively. The updated training dataset will update the mapping between design and response space at each step. This includes correlations of the EM simulation as well as correlation of the desired optical response and the corresponding designs generated. After each cycle the DL models update their weights during each training and learn these correlations more effectively.
	
	\section{Conclusion}
	
	In conclusion, our proposed framework in this paper is capable of achieving rapid and accurate inverse design of metasurfaces for a user-defined optical response with 0.021 MSE and 0.968 cosine similarity. Our framework addresses the design and optimization of metasurface as probabilistic distribution of design space and directs good-quality generation of structural design; simultaneously performing inverse and forward design using DL models. The pGA algorithm facilitates the selection of optimized generated designs for desired optical response to create desired optical response and design space. The generation of desired optical response and  design space alleviates the need for extensive data collection through EM Maxwell's equations solvers and drives our models towards greater generalization including the prediction of new structural designs. The generation of new designs demonstrates that our approach can be generalized to the design of other types of metasurfaces and functionalities.  In addition, this work may be expanded to generate additional data space for other DL-based optimization applications. The results of our deep learning models indicate that this type of framework is a powerful tool to reduce the cost of computation and optimize nanophotonics design efficiency while exploring new designs.

	\newpage
	\section*{References}
	\bibliography{reference.bib}
	
	
	\newpage
	\begin{figure}[h!]
		
		\includegraphics[width=17cm]{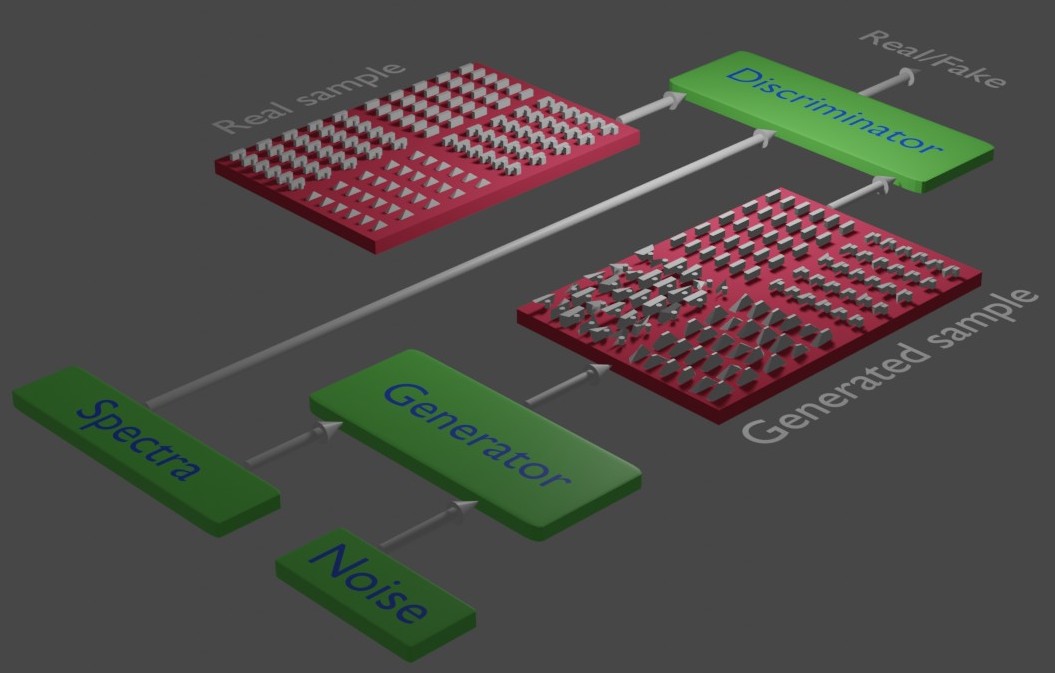} 
	\end{figure}

	
	\newpage
	\begin{figure}[h!]
		
		\includegraphics[width=17cm]{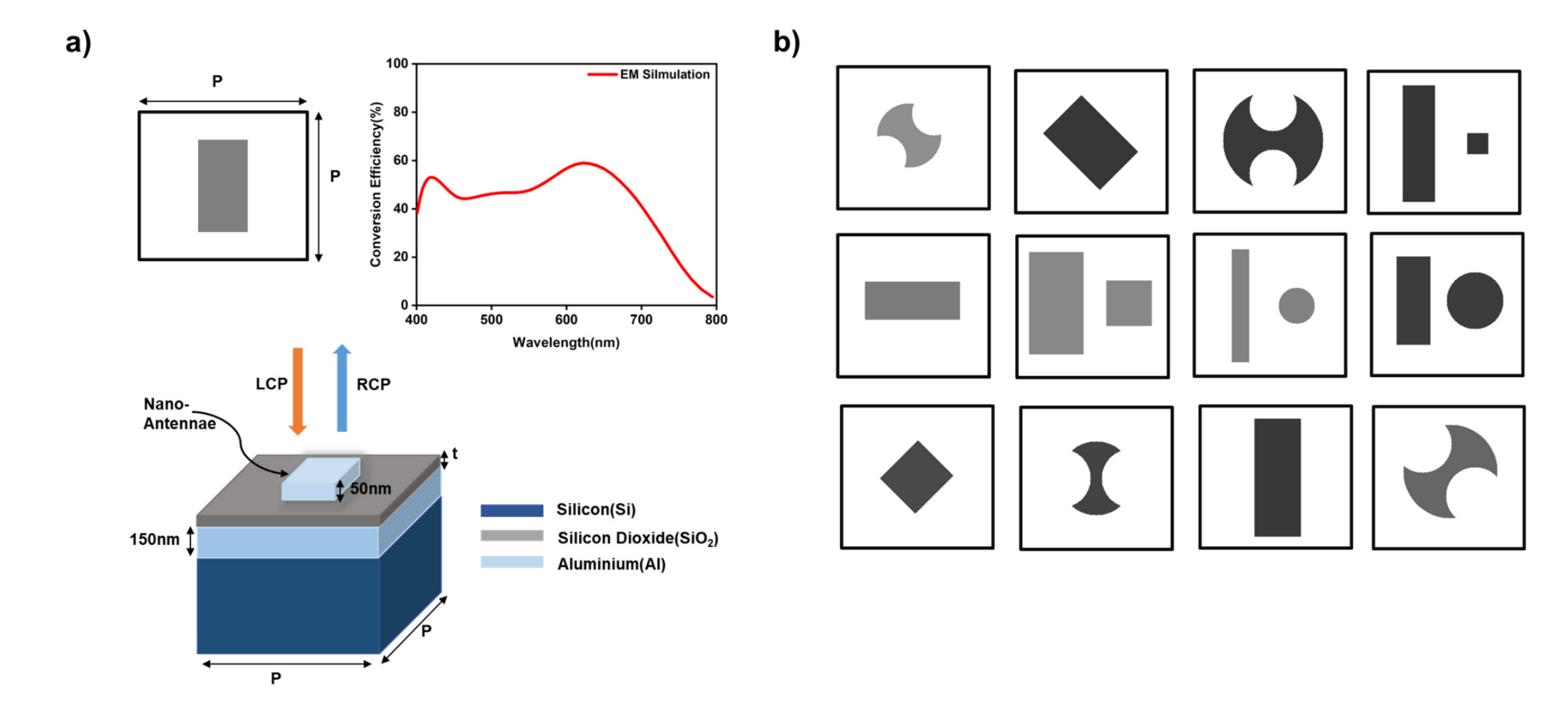} 
		\caption{Illustration of gap-plasmon based HM. (a) A data sample composed of 2D cross sectional image of structural design of Al-nanoantennae and corresponding conversion efficiency as optical response, (b) Examples from structural design classes used for training of DL-NNs : cGAN and SNN. The pixel intensity of structural designs represents the thickness of the SiO$_2$ dielectric spacer.}
	\end{figure}

	\newpage
	\begin{figure}[h!]
		\includegraphics[width=14cm]{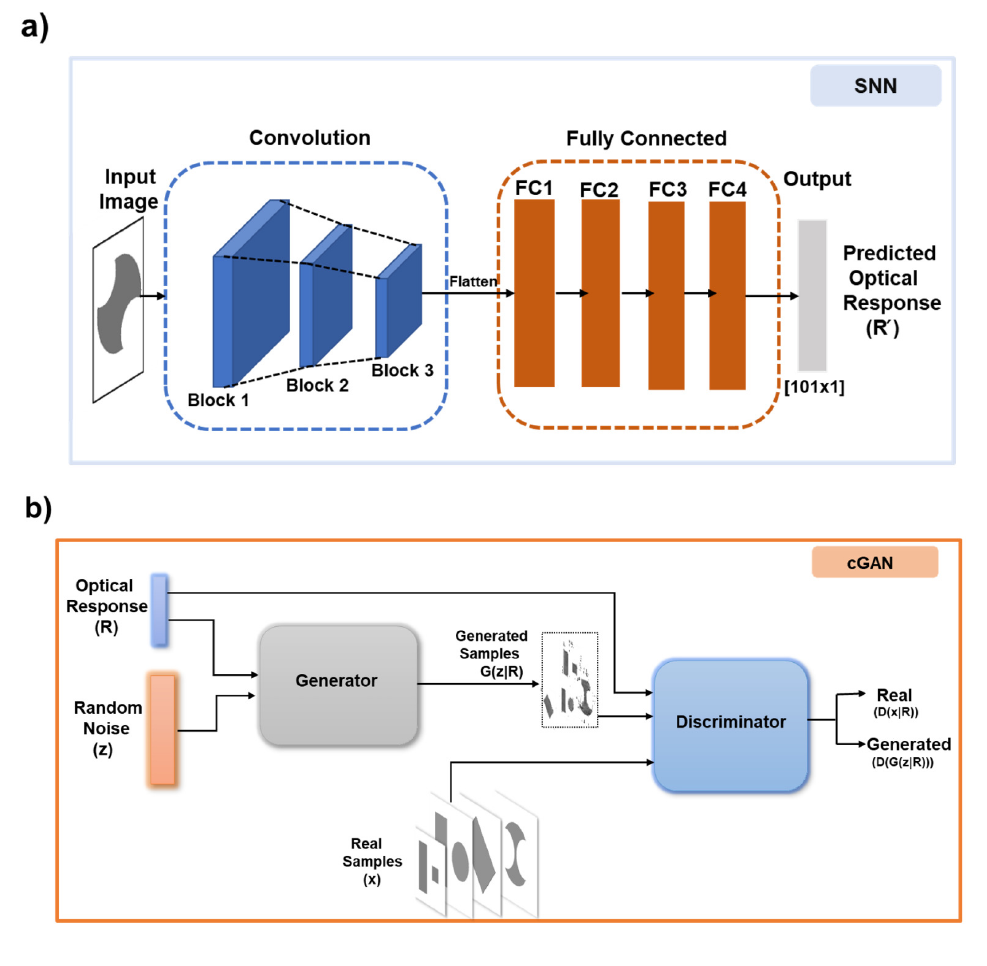} 
		\caption{Schematic illustration of  (a) SNN model architecture composed of CNNs and FC-NNs. It takes 2D cross sectional structural design image as input and predicts it's optical response and (b) cGAN model architecture to generate structural designs. The complete architecture consists of two networks: a generator ($G$) and a discriminator ($D$). $G$ accepts optical response and a random noise distribution $z$ to generate structural design. $D$ evaluates real structural designs from generated structural designs as real or fake.  }
	\end{figure}

	\newpage
	\begin{figure}[h!]
		\includegraphics[width=16cm]{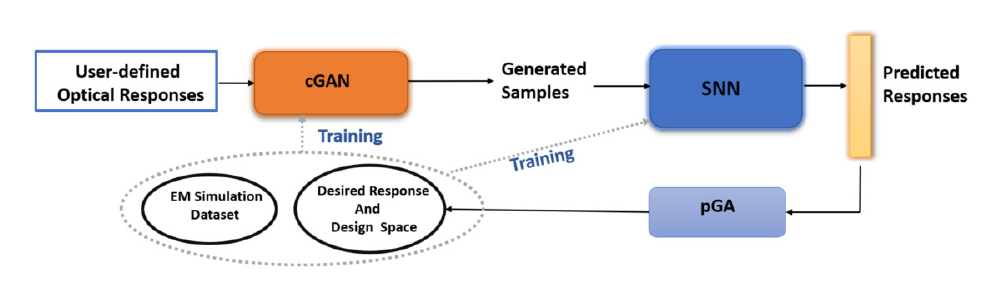} 
		\caption{Schematic of Pseudo Generation Framework. The framework takes as input a batch of user-defined optical responses and generates good-quality structural designs with authentic optical response using cGAN and SNN respectively, and is characterized into desired response and design space via pGA. In each cycle, the EM simulation and new generated desired response and design space are used as an updated dataset for training.}
	\end{figure}

	\newpage
	\begin{figure}[h!]
		\includegraphics[width=16cm]{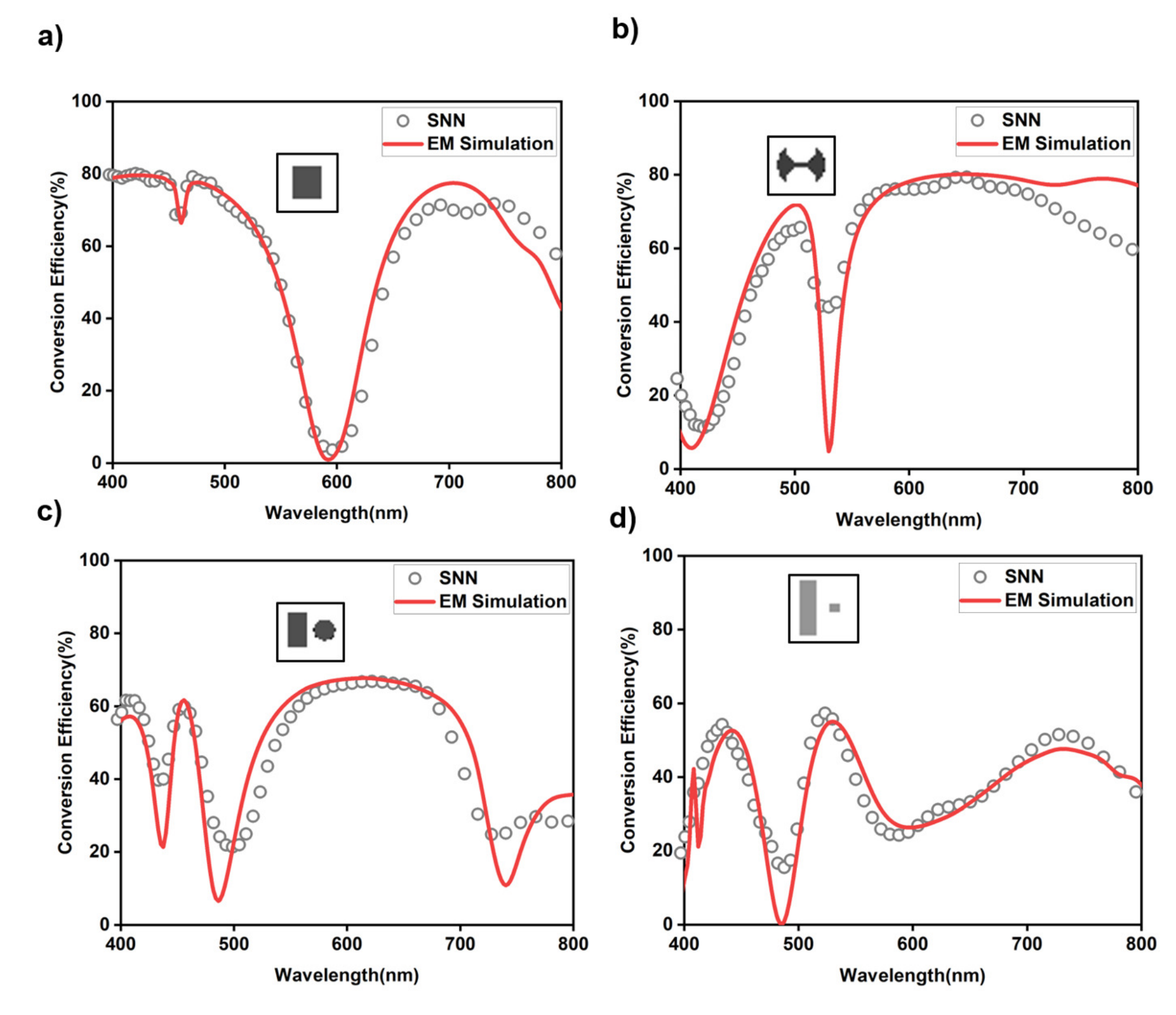} 
		\caption{Test sample examples of SNN predictions (grey circles) with EM simulation spectrum (red curve) on structural design classes, (a) rectangle (b) double-arc (c) rectangle-circle pair (d) rectangle-square pair.}
	\end{figure}

	
	\newpage
	\begin{figure}[h!]
		\includegraphics[width=17cm]{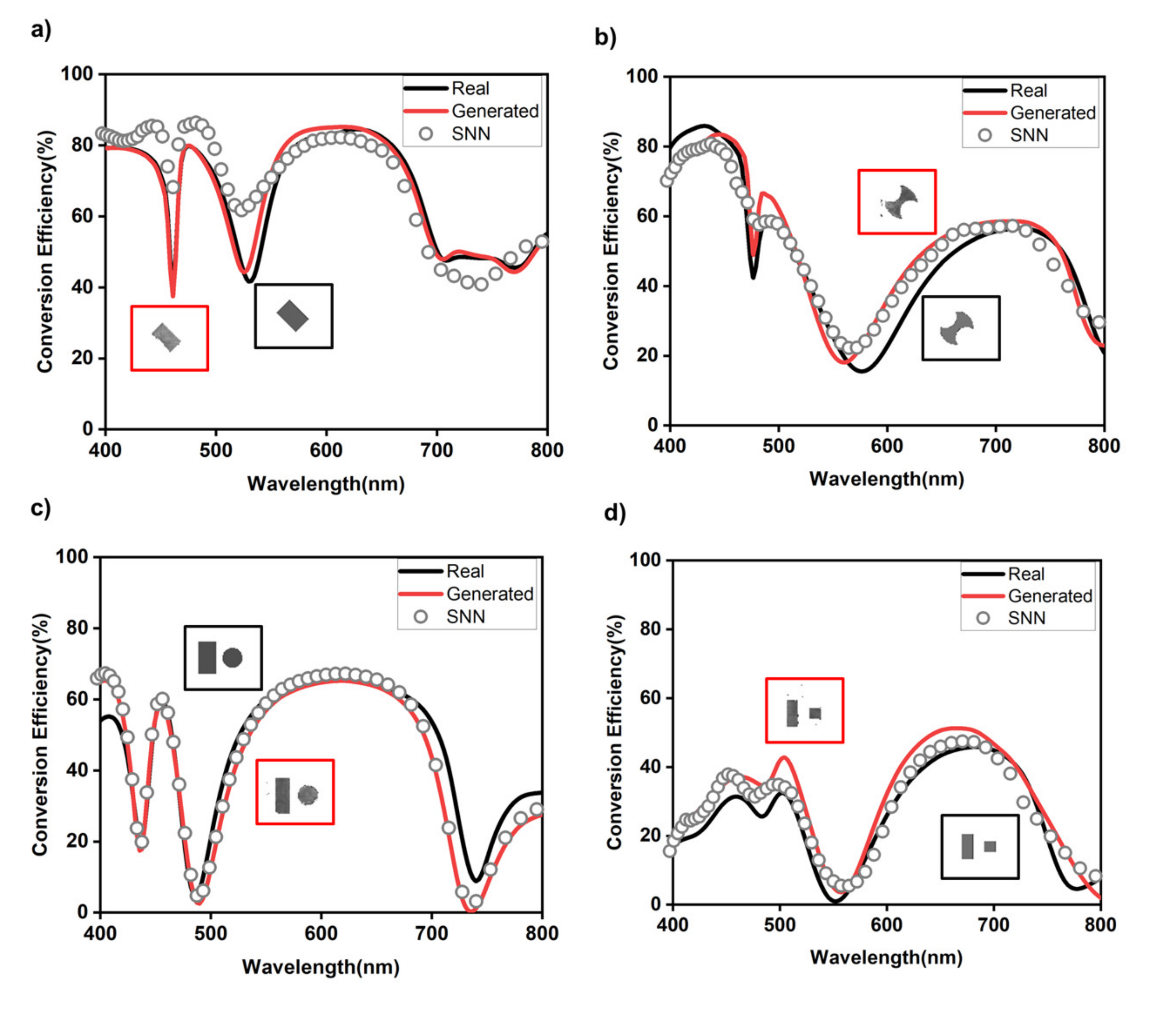} 
		\caption{Evaluation of cGAN on test samples for each class of structural design. The spectrum of EM simulation spectrum for real structural design (black curve) and generated structural design (red curve) along with prediction of SNN for generated structural designs (gray circles) demonstrate good quantitative agreement. Red box inset: Generated structural design, Black box inset: Real structural design.}
	\end{figure}

	
	\newpage
	\begin{figure}[h!]
		\includegraphics[width=15cm]{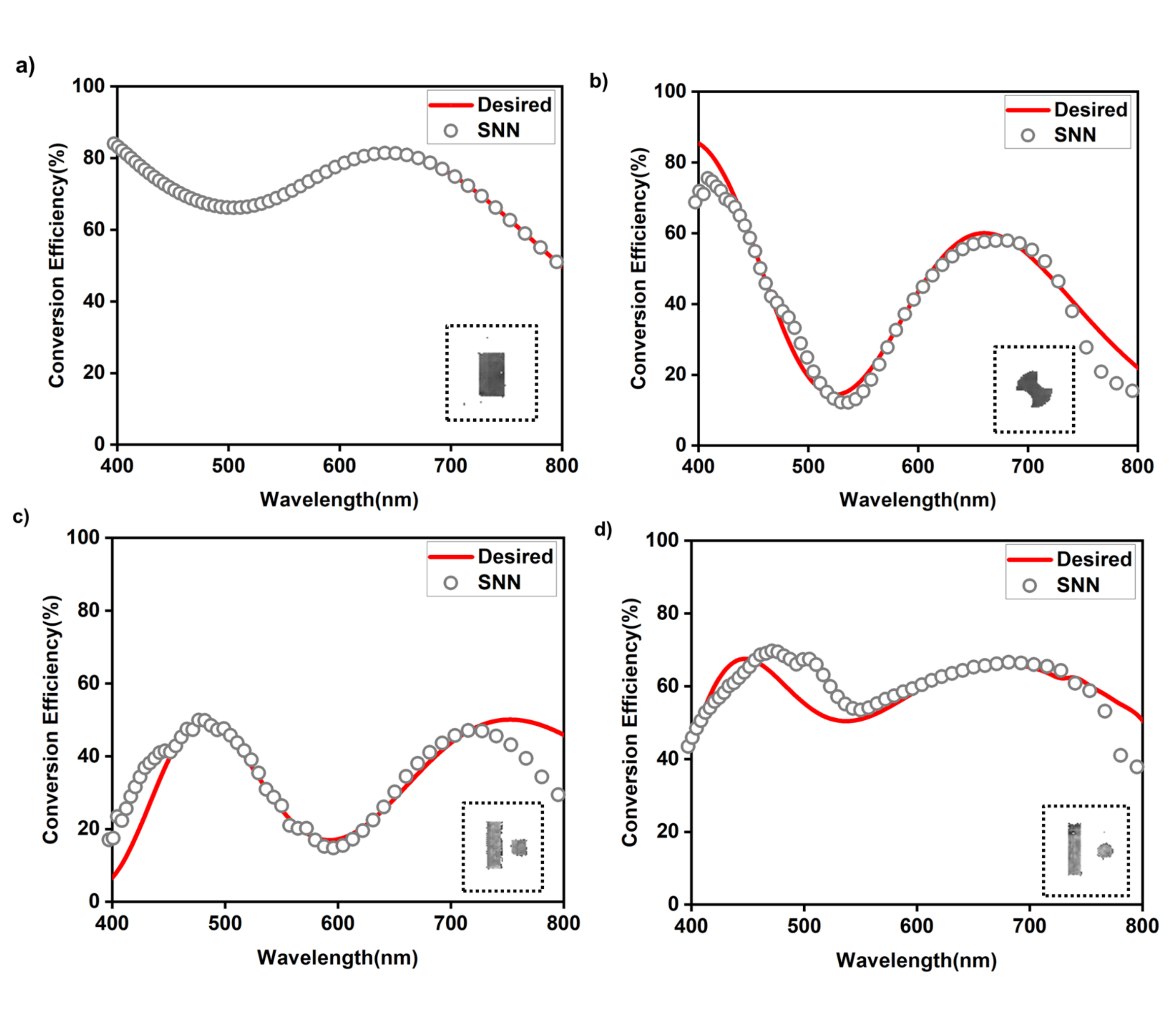} 
		\caption{Examples of structural designs generated for the desired optical response as a Gaussian mixture. The optical response (grey circles) predicted by SNN shows good agreement with  desired optical response (red curve) for generated designs(insets). (a)-(d) generated design for different structural design classes.}
	\end{figure}

	
	\newpage
	\begin{figure}[h!]
		\includegraphics[width=14cm]{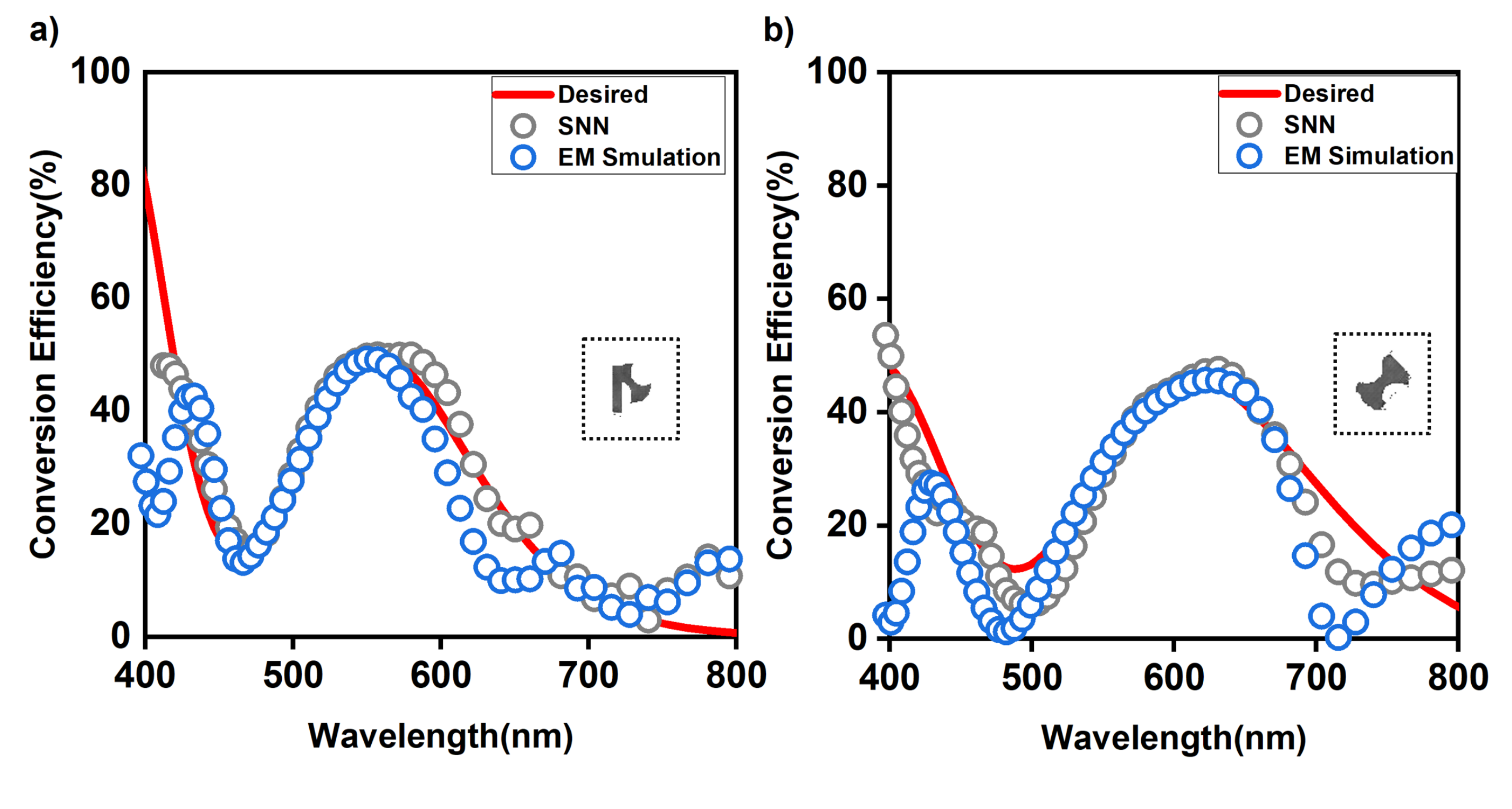} 
		\caption{Examples of new structural designs generated for the desired optical response as a Gaussian mixture. The optical response (grey circles) predicted by SNN and EM Simulation optical response (blue circles) shows good agreement with  desired optical response (red curve) for generated designs(insets). (a) an axe and (b) dumbbell as new generated designs.}
	\end{figure}

	
	\newpage
	
	\section*{Supplementary Information}
	
	\subsection*{I. Variation in optical response of training samples}

	Figure 1 depicts randomly selected data samples from the dataset used for training DL models. The optical responses corresponding to each data sample include sharply varying features such as dips / peak (red arrow), oscillations (orange arrow), and flat (blue arrow) reflections. The appearance in every optical response of a combination of these features induces difficulty in mapping structural design to optical response. The dataset with fewer data samples with these variations could result in deviation in predicting optical response when a NN is being trained on this dataset. Those data samples may act as outliers with spectral response variation. The MSE loss function penalizes outliers in dataset while cosine similarity ensures precise feature prediction.

	
	\centering\includegraphics[width=16cm]{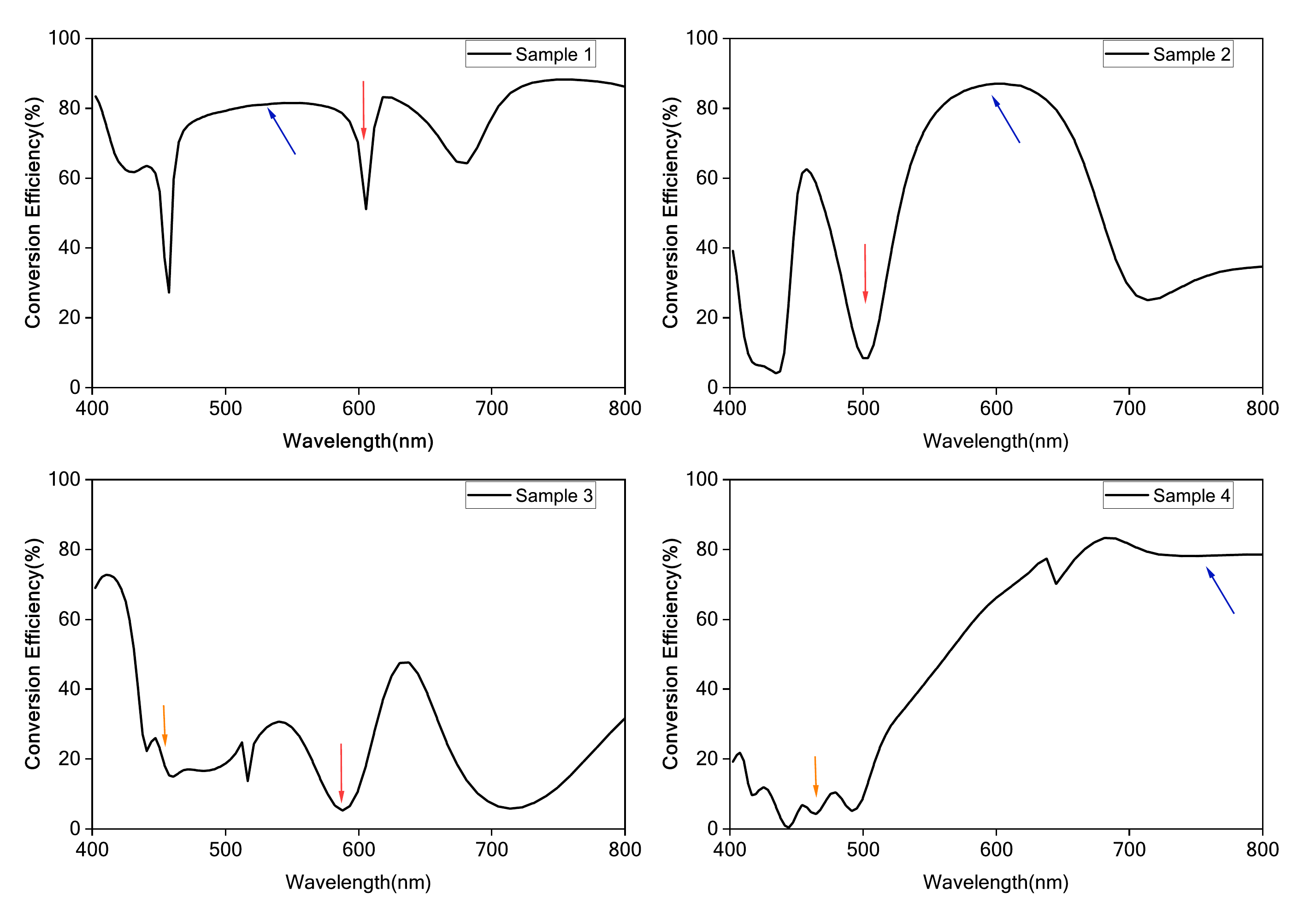}
	\textbf{Figure 1 : Data samples of optical response from EM simulation generated training dataset}

	\newpage
	\begin{table}
		\caption{\textbf{Detailed information of simulation neural network(SNN) model architecture and parameters}}
		
		\begin{tabular}{ |p{3cm}||p{3cm}|p{3cm}|p{3cm}|  }
			\hline
			\multicolumn{4}{|c|}{ \textbf{Convolutional Layers}} \\
			\hline
			\textbf{Layer type} & \textbf{Block 1} & \textbf{Block 2} & \textbf{Block 3}\\
			\hline
			\textbf{Conv2d}  & [1, 32, 3, 1]   & [64, 128, 3, 1] &  [128, 256, 3, 1]\\
			\textbf{Batch Norm2d}  &   32  &  128   &  256\\
			\textbf{ReLU} & &  &  \\
			\textbf{Conv2d}    & [32, 64, 3, 1] & [128, 128, 3, 1 ]&  [256, 256, 3, 1] \\
			\textbf{Batch Norm2d}  &   64  &  128   &  256\\
			\textbf{ ReLU} &   &  &  \\
			\textbf{Maxpool2d} &  (2,2) & (2,2)  & (2,2)\\
			\textbf{Dropout2d} & p = 0.5 & p = 0.5 &  p = 0.5\\
			\hline
			
		\end{tabular}
	\end{table}
	
	\begin{table}
		\begin{tabular}{ |p{3cm}||p{3cm}|p{3cm}|p{3cm}|p{3cm}|}
			\hline
			\multicolumn{5}{|c|}{\textbf{Fully connected Layers}} \\
			\hline
			\textbf{Layer type} & \textbf{FC 1} & \textbf{FC 2} & \textbf{FC 3} & \textbf{FC4}\\
			\hline
			\textbf{Linear}  & 4096  & 2048 &  1024 & 512 \\
			\textbf{Dropout}  &  -   &  p = 0.5   &p = 0.5  & p = 0.5  \\
			\textbf{ReLU} &   -  &  &  &\\
			
			\hline
		\end{tabular}
	\end{table}

	\newpage
	
	\begin{table}
		\caption{\textbf{Network structure and parameters for cGAN model}}
		\begin{tabular}{ |p{3.5cm}|p{3cm}||p{3.5cm}|p{3cm}|  }
			\hline
			\textbf{Layer type} & \textbf{Generator} & \textbf{Layer type} & \textbf{Discriminator}\\
			\hline
			\textbf{ConvTranspose2d}  & [512, 512, 4, 1, 0]   & \textbf{Conv2d} &  [1, 64, 4, 2, 1]\\
			\textbf{Batch Norm2d}  &   512  &  \textbf{Batch Norm2d}   &  64\\
			\textbf{ReLU} & & \textbf{LeakyReLU(0.2)}  &  \\
			\textbf{ConvTranspose2d}    & [512, 256, 4, 2, 1] & \textbf{Conv2d}& [64, 128, 4, 2, 1] \\
			\textbf{Batch Norm2d}  &   256  &  \textbf{Batch Norm2d}   &  128\\
			\textbf{ReLU} &   & \textbf{LeakyReLU(0.2)}  &  \\
			\textbf{ConvTranspose2d}    & [256, 128, 4, 2, 1] & \textbf{Conv2d}& [128, 256, 4, 2, 1] \\
			\textbf{Batch Norm2d}  &   128  &  \textbf{Batch Norm2d}  &  256\\
			\textbf{ReLU} &   & \textbf{LeakyReLU(0.2)} &  \\
			\textbf{ConvTranspose2d}    & [128, 64, 4, 2, 1] & \textbf{Conv2d}&  [256, 512, 4, 2, 1] \\
			\textbf{Batch Norm2d}  &   64  &  \textbf{Batch Norm2d}  &  512\\
			\textbf{ReLU} &   & \textbf{LeakyReLU(0.2)} &  \\
			\textbf{ConvTranspose2d}    & [64, 1, 4, 2, 1] &  \textbf{Conv2d}& [512, 100, 4, 2, 1] \ \\
			
			\textbf{Tanh} &   &  \textbf{Flatten}   & concatenate 101$\times$1 spectrum  \\
			
			&   &  \textbf{FC}   & 512  \\
			&    &  \textbf{Batch Norm1d}  & 512 \\
			&   & \textbf{LeakyReLU(0.2)} &  \\
			&   &  \textbf{FC}   & 1\\
			&   &  \textbf{Sigmoid}   & \\
			\hline
		\end{tabular}
	\end{table}

	\newpage
	
	\begin{table}
		\caption{\textbf{Hyper-parameters for training of cGAN and SNN}}
		\begin{tabular}{ |p{4cm}||p{3cm}|p{3cm}|p{2.5cm} | }
			\hline
			\textbf{Hyper-Parameters} & \textbf{Generator} & \textbf{Discriminator} & \textbf{SNN}\\
			\hline
			\textbf{Batch Size}  & 64   & 64 &  64\\
			\textbf{Learning Rate}  &   2e-4  &  2e-4   &  3e-4\\
			\textbf{Optimizer} & Adam & Adam  & Adam \\
			
			\hline
		\end{tabular}
	\end{table}

	where [input channels, output channels, kernel size, padding]  and  (kernel size, stride) \\
	
	\textbf{FC} = Fully connected layer, \textbf{ConvTranspose} =  Convolutional Transpose Layer, \textbf{Conv} = Convolutional Layer and \textbf{Batch Norm} = Batch Normalization

\end{document}